\documentclass[aps,prb,floatfix,showpacs,twocolumn]{revtex4}
\usepackage{amsmath}
\usepackage{graphicx}

\begin{document}

\title{Influence of solvent quality on polymer solutions: a Monte
  Carlo study of bulk and interfacial properties}
 \author{C.I. Addison, A. A. Louis\footnote{Author for correspondence:
  aal20@cam.ac.uk}, and J.P. Hansen} 
\affiliation{Dept. of Chemistry, University of Cambridge,
Lensfield Road, CB2 1EW, Cambridge, UK}
\begin{abstract}
The effect of solvent quality on dilute and semi-dilute regimes of
polymers in solution is studied by means of Monte Carlo simulations.
The equation of state, adsorptions near a hard wall, wall-polymer
surface tension and effective depletion potentials are all calculated
as a function of concentration and solvent quality.  We find important
differences between polymers in good and theta solvents.  In
the dilute regime, the physical properties for polymers in a theta
solvent closely resemble those of ideal polymers.  In the semi-dilute
regime, however, significant differences are found.
\end{abstract}

\maketitle

\section[Intro]{Introduction}\label{intro}

The properties of polymers in solution are determined by the balance
between effective monomer-monomer attractions and excluded volume
repulsions. Upon increasing the relative strength of the attractions,
for example by cooling, several regimes are encountered.
 As long as the thermal energy $k_B T$ far
exceeds the absolute value $\epsilon$ of the effective monomer-monomer
attraction, good solvent conditions prevail, and individual
polymer coils are swollen due to the dominance of excluded volume
effects. The polymer radius of gyration $R_g$ scales like $L^{\nu}$, where
$L$ is the number of monomers or (Kuhn) segments, and $\nu \approx 0.59$ is
the Flory exponent\cite{deGe79,Rubi03}.  When the temperature is
lowered, the coils shrink due to the action of effective
(solvent-induced) monomer-monomer attraction, until the theta
regime is reached; at the theta-temperature $T_\theta$,
monomer-monomer repulsion and attraction cancel, at least at the
two-body level, and individual coils behave essentially like ideal
(random walk) polymers, such that their radius of gyration scales like
$L^{\frac12}$.  Below $T_\theta$ individual coils collapse into dense
globules with $R_g \sim L^{\frac13}$.  

Moving away from the infinite dilution limit, interactions between
different polymer coils come into play, and give rise to the lowest
order correction to the van t'Hoff limit of the osmotic equation of
state, valid for non-interacting polymers (see Eq.~(\ref{eqHoff}) below).  The
correction, due to pair interactions, is proportional to the square of
the overall monomer concentration $c$; the coefficient is the second
virial coefficient $B_2(T;L)$.  The Boyle temperature $T_B$ is the
temperature at which $B_2$ vanishes, i.e.\ $B_2(T_B,L)=0$.  Since this
condition is met when the monomer-monomer repulsion and attraction
cancel, one expects that $T_B \simeq T_\theta$.

Finally, upon lowering the temperature at a finite polymer
concentration, the polymer solution is found to separate into
polymer-poor and polymer-rich phases below a critical temperature
$T_c$.  Within Flory-Huggins mean-field theory $T_c$ is found to
coincide with $T_B$ within corrections of order $1/\sqrt{L}$.  Thus, a
polymer solution is characterized by three temperatures $T_\theta$,
$T_B$, and $T_c$, of which the first is a single polymer property,
defined in the $L \rightarrow \infty$ limit.  For any given $L$, it
is believed that $T_B(L) > T_\theta > T_c(L)$\cite{Frau97,Yan00}, but
in the scaling limit $L \rightarrow \infty$, $T_B = T_c = T_\theta$.
In fact, when simulating simple models of non-ideal polymers, probably
the most accurate estimate for $T_\theta$ in the scaling limit is
obtained by extrapolating results for $T_B(L)$ to
$1/L=0$\cite{Gras95}.  In this paper the operational definition of
$T_\theta$ will be $T_\theta = T_B$ for sufficiently long chains $(L
\geq 500)$.

\begin{figure}
\includegraphics[width=8cm]{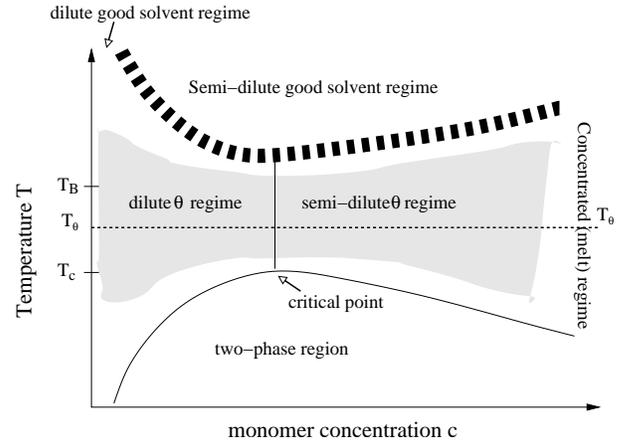}
\caption{ \label{fig:schematic} Schematic phase-diagram of a polymer
solution as a function of temperature $T$ and monomer concentration
$c$, for a finite length $L$.  $R_g$ decreases with decreasing $T$ so
that for good solvents, the monomer concentration at which the dilute
regime crosses over to the semi-dilute regime increases.  As $T$ is
lowered further, there is also a crossover to the ``theta regime'',
denoted schematically by the shaded region.  In the limit $L
\rightarrow \infty$, $T_B$ and $T_c$ both approach $T_\theta$, so that
the size of the ``theta regime'' decreases.  Furthermore, the monomer
concentration at the crossover between the dilute and semi-dilute
regimes, as well as that of the critical point, both tend to zero for
infinite $L$.  }
\end{figure}

For a finite $L$, there is a ``theta solvent regime'' of temperatures
around $T_\theta$, where the polymer behavior most resembles that
predicted for theta polymers.  The longer the polymer, the sharper the
transition between ``theta solvent'' and ``good solvent''
regimes\cite{deGe79,Rubi03,Gras95}.  A typical phase-diagram for
solutions of polymers of finite length is presented in
Fig.~\ref{fig:schematic}.

In the low concentration limit, the physical properties of interacting
polymers in solution closely resemble those of ideal polymers with the
same $R_g$.  For good solvents, however, important deviations from
ideal behavior rapidly set in for increasing concentration, even in
the dilute regime, see e.g.~ref.~\cite{Loui02a}.  On the other hand,
the fact that $B_2(T,L) \simeq 0$ for $T$ near $T_\theta$ suggests
that in the theta regime the osmotic pressure $\Pi$ will follow  van
t'Hoff's law
\begin{equation}\label{eqHoff}
\beta \Pi \simeq \frac{c}{L}
\end{equation}
for a larger concentration range than is found for polymers in good
solvent.  Nevertheless, as concentration is increased further, the
effect of higher order virial coefficients will eventually kick in,
leading to deviations from the simple van t'Hoff
behavior.  Similar differences with ideal-polymer behavior should
also be observable for interfacial properties such as the adsorption
or surface tension near a hard non-adsorbing wall.

While much theoretical and numerical work on simple models of
interacting polymer solutions has been devoted to behavior for
athermal conditions, or at $T=T_\theta$, less is known about how
properties of polymer solutions vary with temperature and
concentration in the intermediate regime $(\infty > T \agt T_\theta)$.

 In the present paper we present MC data for the bulk and interfacial
properties of dilute and semi-dilute polymer solutions for the same
model as used in ref\cite{Gras95}, over a range of temperatures
between the athermal (high temperature) limit and $T_\theta$.  The
main objective is a systematic investigation of polymer solution
properties over a wide range of concentrations, to determine how
equilibrium properties like the osmotic pressure, the adsorption near
a wall, the surface tension and the polymer induced depletion
potential between two walls vary as thermal conditions change from the
good solvent to the theta-solvent regime.  In particular we are
examining the important problem of how the behavior of polymers in
theta-solvent differs from that of ideal (non-interacting) polymers
as their concentration increases.

The paper is organized as follows.  The polymer lattice model and the
computational methodology are introduced in section II.  Results for
the osmotic equation of state and semi-dilute regimes will be
presented and discussed in section III.  We shall next turn our
attention to monomer density profiles near a non adsorbing wall as
well as the polymer-wall surface surface tension as a function of
temperature and concentration (section IV).  The polymer-induced
depletion potential between two walls will be discussed in section VI,
and concluding remarks will be made in section VII.

\section{Model and simulation methodology}\label{simulation}

\subsection{Lattice model of polymer solutions}

A familiar coarse-grained representation of linear polymers in a
solvent, which captures the essential physical features, is a
self-avoiding walk (SAW) lattice polymer, where each lattice site can
be occupied by at most one monomer (or polymer segment) to account for
excluded volume, and where pairs of non-sequential monomers of
nearest-neighbor (n.n.) sites experience an attractive energy $-
\beta \epsilon$.  The parameter $\epsilon$ accounts for the difference
between solvent-solvent, solvent-monomer, and monomer-monomer
interactions, and includes both energetic and entropic
components\cite{deGe79,Rubi03}.  As the dimensionless ratio $\beta
\epsilon = \epsilon/k_B T$ increases from zero (corresponding to the
athermal solvent limit) to larger values, the quality of the solvent
decreases, i.e.\ effective monomer-monomer attractions become
increasingly important, so that the polymer coils tend to shrink until
the theta-regime is reached.  Upon further cooling at finite
concentrations, phase separation sets in at some $T < T_\theta$, see
e.g.\ Fig.~\ref{fig:schematic}.

In this paper we consider polymer chains made up of $L$ monomers in a
simple cubic lattice of $M$ sites (coordination number $z=6$), with
periodic boundary conditions.  If $N$ polymers live on that lattice,
the polymer concentration is $\rho=N/M$, while the monomer
concentration is $c=NL/M$.  $R_g \sim L^\nu$, (with $\nu =0.59$ in
good solvent, and $\nu=\frac12$ in theta solvent) is the radius of
gyration of a polymer coil, and the overlap concentration is defined
as $1/\rho* = \frac43 \pi R_g^3$, where $R_g$, which depends on $\beta
\epsilon$, will conventionally be chosen to be the radius of gyration
in the infinite dilution limit ($\rho \rightarrow 0$) limit.  $\rho*$
separates the dilute regime of the polymer solution ($\rho/\rho* < 1$)
from the semi-dilute regime $(\rho/\rho* > 1)$.

 In the scaling limit $(L \rightarrow \infty)$, the properties of a
polymer solution in the dilute and semi-dilute regimes depend only on
$\rho/\rho*$ and $R_g$, and are independent of the monomer
concentration $c$\cite{deGe79,Rubi03}.  In other words, simulations
with different $c$ but the same $\rho/\rho*$ should give 
equivalent results when length is expressed in  units of $R_g$.  This
requires that  simulations  be carried out
with sufficiently long polymers so that the monomer concentration $c
\ll 1$.  Under good solvent conditions, $R_g
\approx 0.4 L^{0.59}$ for polymers on a simple cubic lattice, so that
\begin{equation}\label{eq1.2a}
  c* = L \rho*  \approx \frac{4}{L^{0.77}},
\end{equation}
whereas in theta-solvent, $R_g \approx 0.55
L^\frac12$\cite{Gras95}, and so
\begin{equation}\label{eq1.0b}
c* \approx \frac{1.4}{L^{0.5}}.
\end{equation}
Most of the subsequent MC simulations have been carried out for
$L=500$ polymers, such that $c* \simeq 0.027$ under good solvent
conditions, while $c* \simeq 0.06$ in theta-solvent.  In order to
ensure that $c < 0.25$ in all simulations, the semi-dilute regime
which could be explored was hence restricted to $\rho/\rho* < 10$ in
good solvent, and $\rho/\rho* < 4$ in theta-solvent.  For higher
concentrations, the c-dependence of the results would no longer be
negligible, and non-universal effects would become important, depending
on the property under consideration.   For this reason, much smaller
polymers, say $L=100$, could not be used to study the semi-dilute
regime.

The previous extensive MC study by Grassberger and Hegger\cite{Gras95}
determined $T_B$ as a function of $L$ for a model identical to the one
used here.  They found that for $L=500$, the Boyle temperature $T_B$
gave $\beta \epsilon=0.265$, the value which we shall use throughout
the discussion in this paper as a reasonable estimate for the
theta-temperature.

\subsection{Monte Carlo simulation methodology}

We used 4 different types of moves to sample configuration space: The
pivot algorithm attempts to rotate part of the polymer around a
randomly chosen segment. Translations move the whole polymer through
space, and reptations remove a segment at one end of the polymer and
attempt to regrow it at the other end.  Since translation and pivot
rapidly become less efficient for increasing polymer concentration, we also
used configurational bias Monte Carlo (CBMC) moves\cite{frenkelbook}.
These are also very effective at lower temperatures, since the
attraction between polymers reduces the accepted number of pivot and
translation moves\cite{Krak03}.

Simulations were carried out in a box of size $160\times100\times100$
for polymers of length $L=500$.  For calculations of radii of
gyration, and other bulk properties, periodic boundary conditions
were used in all directions, whereas for surface properties
calculations, hard walls were placed at $0$ and $160$ along the
x-axis.  The osmotic pressures were calculated using the method of
Dickman\cite{Dick87}.  A external repulsive potential  $\epsilon_w$
that acts between on monomers next to the wall is introduced, and
varied such that the parameter 
$0 \leq \lambda = e^{- \epsilon_w/k_BT} \leq 1$.  $\lambda$=0 prevents
any particles from being adjacent to the wall, and the varying wall
hardness allows the  the osmotic pressure
$\beta \Pi$ to be calculated  from the integral
\begin{equation}\label{eq:Dickman}
\beta \Pi(\rho) = \int_0^1 \frac{d \lambda}{\lambda} \rho_w(\lambda),
\end{equation}
where $\rho_w(\lambda)$ is the site occupation fraction of monomers
right next to the wall. We performed $5$ simulations at different
$\lambda$ values, chosen from a standard Gauss-Legendre distribution,
to carry out the integral above. By using 2 walls of varying hardness,
the statistics are enhanced over using a single wall.  Adding
repulsive walls increases the bulk density at the center of the
simulation cell, an effect which becomes more pronounced for smaller
boxes. Thus $N/V$ should not be taken as the bulk density.  Instead,
we have determined the bulk density by averaging over the region, near
the center of the box, where the influence of the two depletion layers
have completely decayed\cite{Stuk02}.

\begin{figure}
\includegraphics[width=8cm]{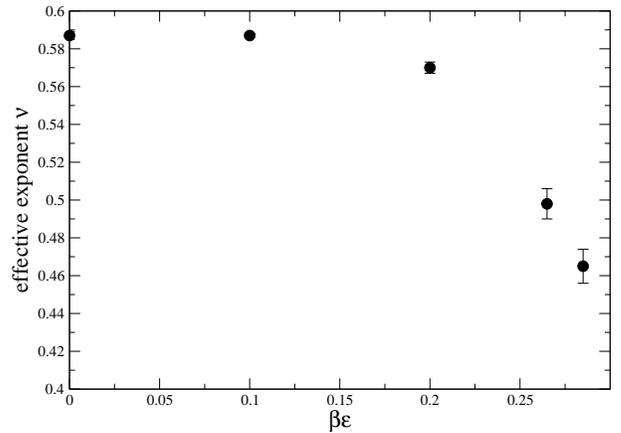}
\caption{ \label{fig:Lvsr} Effective exponents $\nu$, extracted
from simulations with different lengths $L$, as a function of $\beta
\epsilon$.
. Note that the value
presented at $\beta \epsilon = 0.256$, corresponding to $T_B$ for
$L=500$,  was calculated  at slightly different
temperatures, corresponding to $T_B$, for each polymer length.}
\end{figure}

As a first application of the MC code, we have computed the radius of
gyration of an isolated polymer for lengths $L=100$ to $L= 5000$, and
extracted an exponent $R_g \sim L^{\nu}$, shown in
Fig.~\ref{fig:Lvsr}. For higher temperatures, $\nu = 0.588$ within
error bars, as expected for the good solvent regime, but for $\beta
\epsilon = 0.2$ the exponent $\nu$ appears to deviate slightly, and we
find $\nu = 0.570 \pm 0.003 $.  In the limit $L \rightarrow \infty$,
this scaling should be independent of temperature, as long as $T >
T_\theta$.  However, here we observe some rounding, most likely due to
finite length crossover effects.  The precise temperature at which
theta-type scaling behavior of $R_g$ sets in is dependent on length.
The longer the polymer, the smaller the range of temperatures over
which transition from the good-solvent to the theta regime
occurs\cite{Gras95,Shaf99}.  Note also that for a given $L$, the $R_g$
for e.g. $\beta \epsilon =0.1$, is smaller than $R_g$ for $\beta
\epsilon =0$. Nevertheless, the scaling with length, $R_g \sim
L^{0.59}$, is still the same for those two temperatures, i.e.\ the polymers
are still in the good solvent regime.  Thus the effect of lowering the
temperature is simply to renormalize the effective step-length.  

For $T=T_B$, with $T_B$ for each length taken from ref.~\cite{Gras95},
we find $\nu = 0.498 \pm 0.008$, as expected for the theta
regime\cite{Log}. As the temperature is lowered further, $\nu$
continues to decrease, and for low enough $T$ the polymer should
collapse to a compact globule state where $\nu = \frac{1}{3}$.  The
beginning of this trend is evident in the plot for $\beta
\epsilon=0.285$.  The larger error bars reflect the more pronounced
finite $L$ effects expected at this temperature.  For short polymers
the transition from extended to collapsed states occurs over a broad
range of temperatures, whereas for longer polymers the collapse
transition is sharper\cite{Gras95}.

\begin{figure}
\includegraphics[width=8cm]{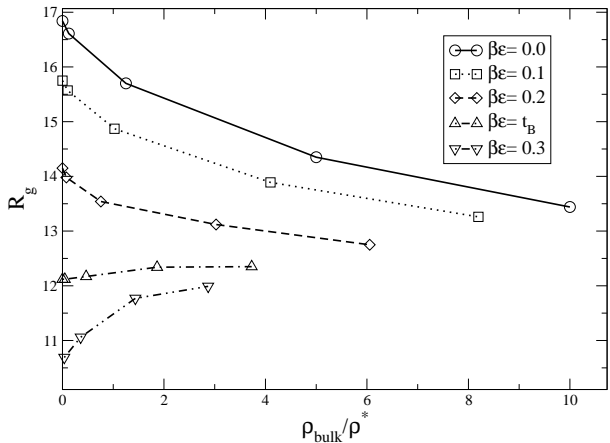}
\caption{ \label{fig:neatrg} $R_g$, calculated for $L=500$ polymers,
decreases with concentration for polymers in good solvent,  is
virtually independent of density for $T=T_B$, and increases with
concentration when $T< T_B$.  For each temperature simulations were
performed up to a maximum monomer concentration $c \approx 0.25$, so
that the total $\rho/\rho*$ range accessible decreases with
temperature. }
\end{figure}

 $R_g$ shrinks (for fixed $L$) as the temperature
is lowered, because the excluded volume interactions are partially
compensated by attractive interactions between monomers.  Similarly
in the good solvent regime, $R_g$ decreases with density because the
excluded volume interactions are screened by overlapping polymer
coils.  This decrease follows a scaling law $R_g \sim
\rho^\frac18$\cite{deGe79,Rubi03} once the semi-dilute regime is well
developed\cite{Bolh01}.  At the theta-temperature, on the other
hand, chain statistics are nearly ideal on all length scales and
concentrations\cite{Rubi03}, so that $R_g$ should be independent of
concentration.  For even lower temperatures, $T_B > T > T_c$, the
screening of attractive interactions now implies that $R_g$ should
increase with concentration.  These trends are indeed observed in our
MC simulations, as illustrated in Fig.~\ref{fig:neatrg} for $L=500$
polymers.

\section{Equation of state in the dilute and semi-dilute regimes}\label{sectionIII}

\begin{figure}
\includegraphics[width=8cm]{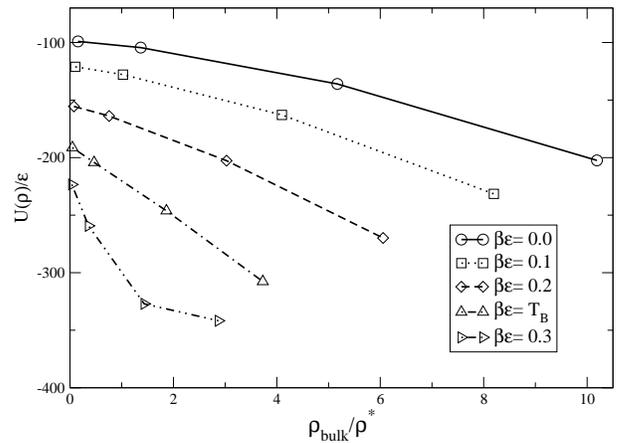}
\caption{ \label{fig:energies} MC data for the reduced internal energy
$U(\rho)/ \epsilon = - \bar{N}_c$ for $L=500$ chains, plotted as a
function of polymer concentration $\rho/\rho*$ for 5 different
temperatures.  Note how the curvature changes with solvent
quality.}
\end{figure}

The thermodynamic property which is most readily extracted from the
simulations is the internal energy of polymer solutions, which is
simply $U=-\bar{N}_c \epsilon$, where $\bar{N}_c$ is the average
number of nearest-neighbor ``contacts'' between non-connected
monomers.  The energy per polymer is $U/N = - \epsilon L \bar{n}_c$,
where $\bar{n}_c$ is the average number of non-connected nearest
neighbors around a monomer.  Within a simple mean-field picture,
$\bar{n}_c$ is expected to increase linearly with the polymer
concentration. Such linear behavior is indeed observed in
Fig.~\ref{fig:energies} for theta conditions.  At higher
temperatures, there is a significant downward curvature on $U(\rho)$,
while below $T_B$, but above $T_c$, $U(\rho)$ curves upward.

\begin{figure}
\includegraphics[width=8cm]{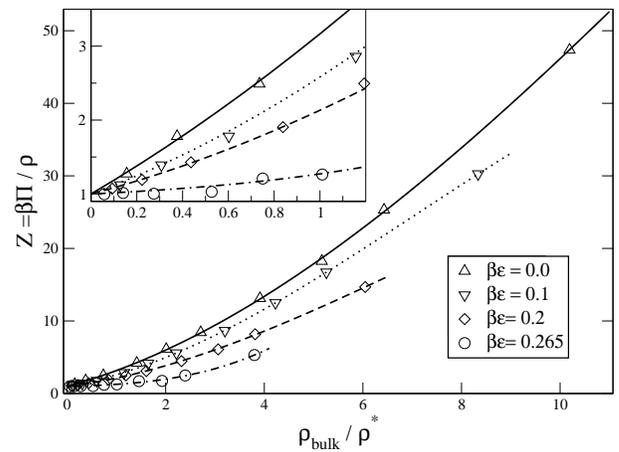}
\caption{ \label{fig:neatopress} Linear graph of the EOS
$Z=\beta \Pi(\rho)/\rho$ as a function of concentration for different
temperatures. (Inset) A blowup of the dilute regime highlights the
differences between near theta solvents and polymers in  good
solvent.  The former show behavior that mimics that
of ideal polymers.}
\end{figure}

\begin{figure}
\includegraphics[width=8cm]{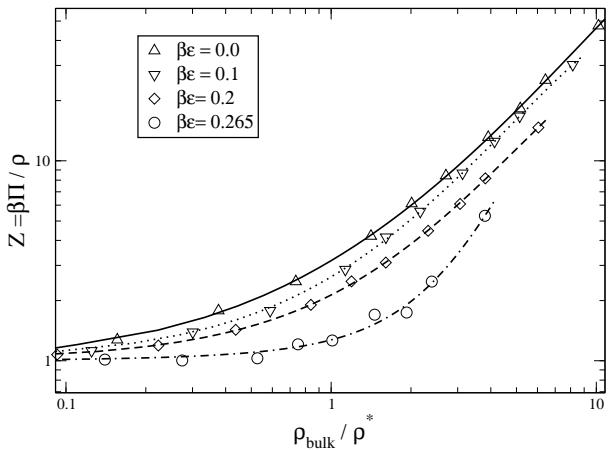}
\caption{ \label{fig:neatlogplot} The EOS as a function of
 concentration for different temperatures, plotted on a double
 logarithmic scale.  In the semi-dilute regime, the EOS is consistent
 with scaling theory, which predicts $\beta \Pi(\rho)/\rho \sim
 \rho^{1.3}$ for good solvents and $\beta \Pi(\rho)/\rho \sim
 \rho^{2}$ for theta solvents.  }
\end{figure}

The osmotic equation of state (EOS) $Z=\Pi(\rho)/(\rho k_B T)$ was
calculated using Dickman's method\cite{Dick87} described in the
previous section.  The MC results for $Z$ as a function of
$\rho/\rho*$, for various temperatures, are plotted in
Figures~\ref{fig:neatopress}~and~\ref{fig:neatlogplot} on linear and
logarithmic scales.  As expected, Z increases with $\rho/\rho*$, and
decreases with  temperature\cite{crossover}. 
 Since the theta-solvent regime is defined by
$T \approx T_B$, the EOS is expected to be very flat at low
polymer concentrations since $T_B$ is defined as the temperature at
which the second virial coefficient $B_2(T,L)$ vanishes in the virial
expansion\cite{deGe79,Rubi03}:
\begin{equation}\label{eos}
Z = \frac{\beta \Pi}{\rho} = 1 + B_2 c + B_3 c^2 + {\cal O}(c^3).
\end{equation}
This is indeed confirmed by the MC results in the dilute regime, as
illustrated in the inset of Fig.~\ref{fig:neatlogplot}.  Up to
$\rho/\rho* \simeq 1$, $Z$ hardly rises with $\rho$, i.e.\ the system
behaves like a solution of ideal polymers.  At higher concentrations,
however, interactions between monomer triplets come into play, so
that, according to Eq.~(\ref{eos}), $Z$ is expected to increase like
$\rho^2$, in agreement with a standard scaling
argument\cite{deGe79,Rubi03}.  The same scaling argument predicts that
the EOS of polymers in good solvent scales as $Z \sim
\rho^{1/(3 \nu -1)} \sim \rho^{1.3}$; the exponent $1.3$
indeed agrees with the slopes of the double logarithmic plots of
$Z(\rho)$ shown in Fig.~\ref{fig:neatlogplot}, for $\rho/\rho* \agt 1$
and $0 \leq \beta \epsilon \leq 0.2$.  The slope of the
theta-temperature EOS is clearly significantly larger, and
compatible with the expected exponent of $2$.  However, for the
$L=500$ chains used in the present simulations, the accessible
semi-dilute regime of concentrations is too small to allow a fully
quantitative estimate (c.f.\ the discussion in section II). Since the
EOS of theta-polymers scales with a larger exponent than that of
polymers in good solvent, there will eventually be a concentration $\rho/\rho*$
where the EOS of the theta-polymers becomes larger than that of
polymers in good solvent.  This may, at first sight, seem
counter-intuitive, but it should be remembered that for a given size
$R_g$, theta-polymers are more compact.  By extrapolating our
simulation data to larger $\rho/\rho*$, we expect the cross-over
between theta-polymers and $\beta\epsilon=0$ (SAW) polymers to occur at
$\rho/\rho* \agt 35$ which is, in practice, very hard to achieve both
in simulations and experiments\cite{extrapolation}.  On the other
hand, if one starts from a solution under good solvent conditions,
i.e.\ ``high'' temperature, at a given $\rho$, and lowers the
temperature, then the EOS will decrease monotonically, because $R_g$,
and with it the ratio $\rho/\rho*$, decreases with temperature.

\section{Monomer density profiles near a non-adsorbing
  wall}\label{sectionIV}

For entropic reasons, polymer solutions will be highly depleted in
the vicinity of a non-adsorbing hard wall: since fewer conformations
are available to a polymer coil the wall will effectively repel the
polymers so that their concentration will drop from the bulk value for
distances $z \agt R_g$, to a much lower value at contact\cite{deGe79}.
The case of polymers in good solvent (in the SAW limit) has been
investigated in much detail in ref.~\cite{Loui02a}, where important
deviations from ideal polymer behavior were found even in the dilute
regime.  Here we continue these investigations, but for polymers in
 solvents of varying quality.

\subsection{Monomer density profiles}

Fig~\ref{fig:neatdprofeps2} compares the present MC results for the
reduced monomer density profiles $c(z)/c$ as functions of the distance
$z$ from the planar surface in the good solvent $(\beta \epsilon =0$)
and theta-solvent ($\beta \epsilon =0.265$) limits for $L=500$
chains, and several bulk concentrations.  The differences between the
two regimes are very significant.  In the dilute regime the density
profiles in the theta solvent hardly change, and are remarkably
close to that of ideal polymers, while the profiles in good solvent
already differ significantly from the latter, and are ``pushed'' closer
toward the hard wall.  In the semi-dilute regime, on the other hand,
the density profiles of polymers in theta-solvent differ
considerably from that of ideal polymers (which is independent of
concentration) and instead more closely resemble those of polymers in
good solvent, although the depletion ``hole'' is systematically wider
for the former compared to that of the latter.  This is illustrated in
Fig.~\ref{fig:neatdprof}, where the profiles at three different
temperatures are compared at low concentration, and in the semi-dilute
regime.

\subsection{Reduced adsorption}

The reduced adsorption of polymers near a hard wall is defined as\cite{Rowl89}:
\begin{equation}\label{eq2.3}
\hat{\Gamma}(\rho) = - \frac{1}{\rho}\frac{\partial (
 \Omega^{ex}/A)}{\partial \mu} = \int_{0}^\infty h(z) dz,
\end{equation}
where $\Omega^{ex}/A$ is the surface excess grand potential per unit
area $A$, $\mu$ is the chemical potential, and $h(z) = c(z)/c -1$.
Because of the depletion hole, $\hat{\Gamma}(\rho)$ is negative, and
has the dimension of length.  One could replace $c(z)$ by the
distribution of the center of mass (CM) of the polymers
$\rho_{cm}(z)$. Although the reduced profiles $h_{cm}(z) =
\rho_{cm}(z)/\rho-1$ would look different from $h(z)$ (see e.g.\ Fig.1
of Ref.~\cite{Loui02a}), the reduced adsorptions would be
identical to those calculated from the monomer density profiles, due
to the conservation of the number of monomers.

\begin{figure}
\includegraphics[width=8cm]{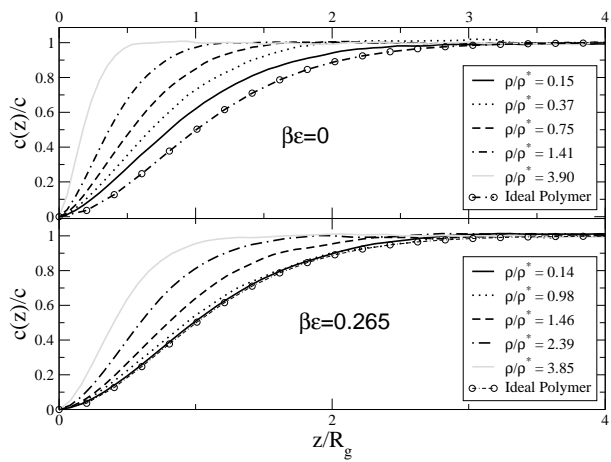}
\caption{\label{fig:neatdprofeps2} Monomer density profiles $c(z)/c$
for polymers near a hard non-adsorbing wall.  The upper panel shows
the density profiles for polymers in good solvent, while the lower
panel is for polymers at $T=T_B$.  At these near theta conditions, the
density profiles in the dilute regime are nearly indistinguishable
from  those of ideal
polymers, but important deviations emerge in the semi-dilute regime.
}
\end{figure}

\begin{figure}
\includegraphics[width=8cm]{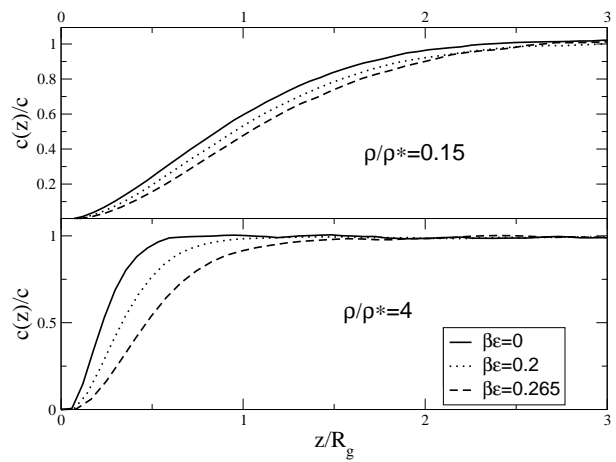}
\caption{ \label{fig:neatdprof} Monomer density profiles as a function
of temperature for two densities.  The upper panel is for $\rho/\rho*
\approx 0.14$, and the lower panel is for $\rho/\rho* \approx 4$, well
into the semi-dilute regime.  The density profiles at lower
temperatures are more extended when plotted in terms of $z/R_g$.  }
\end{figure}

For ideal polymers, the reduced adsorption can be exactly calculated
to be\cite{Asak54,Bolh01}:
\begin{equation}\label{eq2.3b}\hat{\Gamma}^{id} = -
 2R_g/\sqrt{\pi} \approx - 1.1284 R_g,
\end{equation}
which is independent of concentration.  For polymers in good solvent,
a renormalization group (RG)\cite{Shaf99} calculation predicts that
$\hat{\Gamma}(0) \simeq -1.074 R_g$ for the adsorption in the
low-density limit\cite{Hank99}, which is close to, but slightly less
negative than the ideal result~(\ref{eq2.3b}).  As the polymer
concentration increases, interactions between monomers of non-ideal
polymers cause the depletion layer to shrink, as illustrated by the
density profiles shown in 
Figs.~\ref{fig:neatdprofeps2}~and ~\ref{fig:neatdprof}.  Thus, in
contrast to the case of ideal polymers, where $c(z)/c$, and therefore
$\hat{\Gamma}$ is independent of concentration, for interacting
polymers the reduced adsorption $\hat{\Gamma}$ should become less
negative for increasing polymer concentration.  This is indeed what is
observed in Fig.~\ref{fig:neatadsorb}, where the variation of the
reduced adsorption with concentration is shown for several temperatures.

\subsubsection{Adsorption in the dilute regime}

\begin{figure}
\includegraphics[width=8cm]{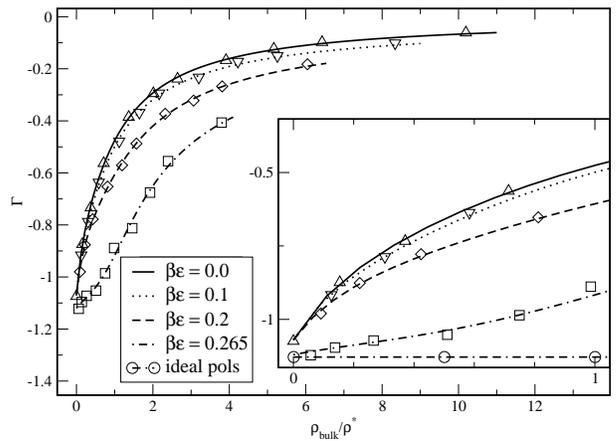}
\caption{\label{fig:neatadsorb} Reduced adsorption $\hat{\Gamma}/R_g$ as a
  function of concentration for several temperatures.  (Inset) Blowup
  of the dilute regime.  Note the different curvature at low
  densities.  The values at $\rho/\rho*=0$ are take from theory to be
  $\hat{\Gamma}/R_g = 1.074$\protect\cite{Hank99} for the first three
  temperatures, and $\hat{\Gamma}/R_g= 2/\sqrt{\pi}$ for $T=T_B$. }
\end{figure}

In the dilute regime $(\rho/\rho* \alt 1)$, there is, once again, a
qualitative difference between polymers in theta or good solvent,
as highlighted in the inset of Fig.~\ref{fig:neatadsorb}.  While the
curvature of the $\hat{\Gamma}(\rho)$ curves with concentration is
negative in good solvent conditions, it is clearly positive under
theta-conditions, and changes to negative in the semi-dilute
regime.  Up to $\rho/\rho* \simeq 1$, the results for $\hat{\Gamma}$
remain close to (but above) the ideal polymer result~(\ref{eq2.3b}).
Note, however, that the relative deviation from ideal polymers is more
pronounced than for the osmotic EOS (compare the insets to
Figs.~\ref{fig:neatopress}~and~\ref{fig:neatadsorb}), hinting at a
stronger effect of three-particle interactions on the adsorption of
theta-polymers.

\subsubsection{Adsorption in the semi-dilute regime}

In the semi-dilute regime we expect the adsorption to be proportional
to the correlation length $\xi/R_g \sim (\rho/\rho*)^{-\nu/(3
\nu-1)}$\cite{deGe79,Rubi03}, since this is the only relevant
length-scale. Note that the numerical prefactors for different
semi-dilute correlation lengths vary with the property one is
attempting to describe, see e.g.~\cite{Huan02} for a very useful
discussion of this matter.  In our previous paper\cite{Loui02a} we
showed that for polymers in good solvent, the reduced adsorption
$\hat{\Gamma}/R_g \sim (\rho/\rho*)^{-0.77}$ is consistent with this
scaling.  Here we present more simulations in the semi-dilute regime
to confirm this behavior, finding that $\hat{\Gamma}/R_g \sim -0.4
(\rho/\rho*)^{-0.77}$ provides a good fit.  Scaling theory
considerations also predict that $\xi/R_g \sim (\rho/\rho*)^{-1}$ for
theta solvents\cite{deGe79,Rubi03}.  However, this behavior cannot be
reliably extracted from the current simulations of $\hat{\Gamma}$,
which are not performed for a large enough range of
$\rho/\rho*$. Furthermore, we expect that for the higher values of
$\rho/\rho*$ sampled here, finite $c$ effects may already be coming
into play.

\section{Wall-polymer surface tension}

\begin{figure}
\includegraphics[width=8cm]{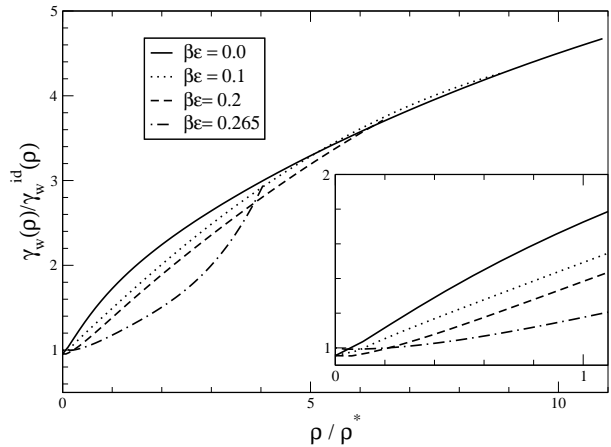}
\caption{\label{fig:neatst3} Wall-polymer surface tension divided by
 the ideal surface tension for various solvent qualities, plotted as
 a function of concentration $\rho/\rho*$.  (Inset) A blowup of the
 dilute regime highlights the differences between
 $\gamma_w/\gamma_w^{id}$ for good and theta solvents.}
\end{figure}

The wall-polymer surface tension, i.e.\ the free-energy cost of
introducing a non-adsorbing hard wall and its associated depletion
layer, can be calculated from the adsorption and EOS by use of the
Gibbs adsorption equation\cite{Mao97}:
\begin{equation}\label{eq2.4}
\gamma_w(\rho) = \frac{ \partial \Omega^{ex}}{\partial A} = -
\int_{0}^{\rho} \left( \frac{\partial \Pi(\rho')}{\partial \rho'}\right)
\hat{\Gamma}(\rho') d\rho'.
\end{equation}
  By performing one  integration by parts
w.r.t.\ density, Eq.~(\ref{eq2.4}) can also be expressed as:
\begin{equation}\label{eq2.4b}
\gamma_w(\rho) = -\Pi(\rho) \hat{\Gamma}(\rho) +
\int_{0}^{\rho} \Pi(\rho')\left( \frac{\partial
\hat{\Gamma}(\rho')}{\partial \rho'}\right) d\rho'.
\end{equation}
  The first term in this equation has an appealing physical
interpretation as the free energy cost per unit area of creating a
slab cavity of width $\hat{\Gamma}(\rho)$.  For ideal polymers, where
$\hat{\Gamma}(\rho)$ is independent of concentration, this term is the
only one that contributes, and so $\gamma_w^{id} = - \rho
\hat{\Gamma}^{id} = -\rho 2/\sqrt{\pi} $.  Since the EOS and
$\hat{\Gamma}(\rho)$ for interacting polymers in various quality solvents
were calculated in the previous sections, we can now, using
Eq.~(\ref{eq2.4b}), determine the surface tension for different
solvent qualities.  Results are shown in Fig.~\ref{fig:neatst3}.  For
good solvent conditions, i.e.\ $\beta \epsilon = 0$, they are, to
within simulation errors, in near quantitative agreement with the RG
calculations of Maasen, Eisenriegler, and Bringer\cite{Maas01}. In
fact, the larger number of simulation results in the semi-dilute
regime, used in the present work, lead to even better agreement than
that shown earlier in Ref.~\cite{Loui02a}.  We are not aware of
theories of comparable quality for the surface tension of theta
polymers.


As discussed earlier for the EOS and the adsorption, we again find
 qualitatively different behavior for the $T=T_B$ solution in the
 dilute regime, as highlighted in the inset of Fig.~\ref{fig:neatst3}.
 The surface tension for such theta like solvents resembles that of
 ideal polymers much more closely than that of polymers in  good
 solvent.  As expected from the related behavior of the adsorption,
 the relative deviation from ideal polymer behavior in the dilute
 regime is somewhat larger than what was found for the EOS.  Note that
 the curves $\gamma_w/\gamma_{id}$ cross at very low
 $\rho/\rho*$. This follows from the fact that the low concentration limit
 of the surface tension is given by $\lim_{\rho \rightarrow 0}
 \beta \gamma_w = - \rho \hat{\Gamma}$, and $-\hat{\Gamma}$ is smaller for
 interacting polymers than for ideal or theta polymers. (See
 Fig.~\ref{fig:neatadsorb}).  At slightly higher densities, the
 curves cross because $\gamma_w$ increases more rapidly with concentration
 for good solvents than for theta solvents.


In the semi-dilute regime, where $\hat{\Gamma} \sim \xi \sim \rho^{-
  \nu/(3 \nu -1)}$ and $\Pi(\rho) \sim \rho^{3 \nu/(3\nu -1)}$,
  Eq.~(\ref{eq2.4}) simplifies\cite{Loui02a} to 
\begin{equation}\label{eq2.5}
\gamma_w^{sd}(\rho) \approx \frac{3}{2} \Pi \hat{\Gamma} \sim \rho^{2
  \nu/(3 \nu -1)}.
\end{equation}
This implies $\gamma_w^{sd} \sim \rho^{1.539}$ semi-dilute scaling for
polymers in good solvent, and $\gamma_w^{sd} \sim \rho^2$ behavior for
theta solvents, which is consistent with the results in
Fig.~\ref{fig:neatst3}.  Furthermore, this scaling suggests that for
large enough $\rho/\rho*$, the surface tension curve for theta
polymers will eventually cross that of polymers in good solvent.
Indeed, Fig.~\ref{fig:neatst3} already appears to suggest this
behavior, although some care must be taken in extracting the
$\rho/\rho*$ at which crossover occurs, due to the the possibility of
finite $c$ effects for theta polymers\cite{extrapolation}. With these
caveats in mind, our results suggest that the surface tension for
theta polymers would be greater than that of interacting polymers when
$\rho/\rho* \geq 10$, which is considerably lower than the predicted
crossing for the EOS, and within reach of experiment and simulations.
On the other hand, just as in the case of the EOS, if one were to
lower the temperature of an experimental polymer solution at a given
$\rho$, this would lead to a lower $\rho/\rho*$, so that a monotonic
behavior of $\gamma_w$ with temperature is expected when following
this route.

\section{Depletion potentials between two walls}

The depletion potential between two walls or plates induced by a
polymer solution is defined as the difference in free energy between
the cases where the plates are at a distance $z$, or infinitely far
apart.
 The polymer solution between the
plates is taken to be in equilibrium with a much larger reservoir of
pure polymer solution at the same chemical potential.  At infinite
distance, the free energy cost of  having two plates is simply twice the
cost of making a depletion layer, while at contact, these two depletion
layers are destroyed.  Thus, when $z=0$, the free energy per unit area
of plate is given by
\begin{equation}\label{eq3.1}
W(0) = - 2 \gamma_w(\rho).
\end{equation}

 For small $z$ virtually no polymer is expected to penetrate between
the plates, which are hence pressed together by the osmotic pressure
$\Pi(\rho)$, leading to a linear increase of $W(z)$ with $z$.  As discussed in
ref~\cite{Loui02b}, the simplest approximation for the depletion
potential  is to continue this linear form 
for larger $z$:
\begin{eqnarray}\label{eq2.6}
W(z)  = & W(0) + \Pi(\rho) z; & \,\,\,\, z \leq D_w(\rho)   \nonumber \\
W(z)  = & \,\,\, 0
\,\,\,\,\,\,\,\,\,\,\,\,\,\,\,\,\,\,\,\,\,\,\,\,\,\,\,\,\,\,; 
 & \,\,\,\, z > D_w(\rho)   
\end{eqnarray}

where the range is given by
\begin{equation}\label{eq2.6a}
D_w(\rho) = -\frac{W(0)}{\Pi(\rho)} = \frac{2 \gamma_w(\rho)}{\Pi(\rho)}.
\end{equation} 
  We note that this approximation is similar to that adopted by
Joanny, Leibler, and de Gennes who, in their pioneering
paper\cite{Joan79}, approximated the force between two plates as
constant for $x \leq \pi \xi(\rho)$ and zero for $x > \pi
\xi(\rho)$. This also results in a linear depletion potential.

In ref~\cite{Loui02b}, we showed that this simple theory is virtually
quantitative for polymers in good solvent\cite{Tuin03a}.  Furthermore,
it is well known to be quite accurate for ideal polymers as
well\cite{Bolh01}.  Although one could easily generalize the theory to
include the small curvature seen for ideal polymers close to
$z=D_w$\cite{Asak54}, we ignore these small corrections in the
interests of simplicity.  Since this theory is accurate for polymers
in good solvent as well as for ideal polymers, we postulate that the
same simple assumptions are valid for other solvent qualities, and
use Eq.~(\ref{eq2.6}) to calculate the depletion potentials.

\begin{figure}
\includegraphics[width=8cm]{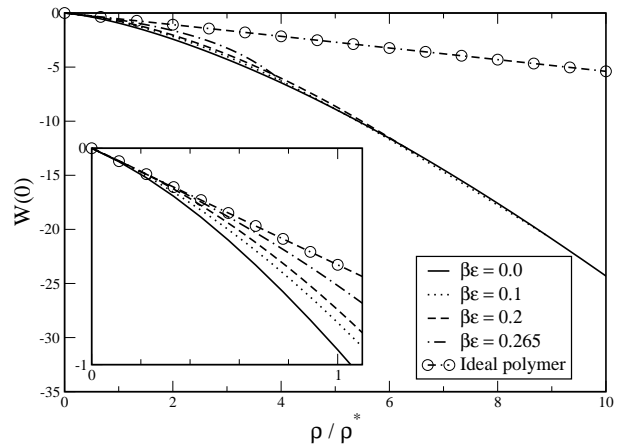}
\caption{\label{fig:neatW0} The depletion potential at contact is
 given by $W(0) = - 2 \gamma_w(\rho)$. (Inset) A blowup of the
 dilute regime shows that  theta polymers most closely resemble
 the ideal polymers.  }
\end{figure}

The well-depth $W(0)$ follows from Eq.~(\ref{eq3.1}), and is shown in
Fig.~\ref{fig:neatW0}.  The behavior mirrors that of
Fig.~\ref{fig:neatst3} of course.  From the inset it is clear that
theta polymers most closely resemble ideal polymers in the dilute
regime, while important deviations are found in the
semi-dilute regime.

\begin{figure}
\includegraphics[width=8cm]{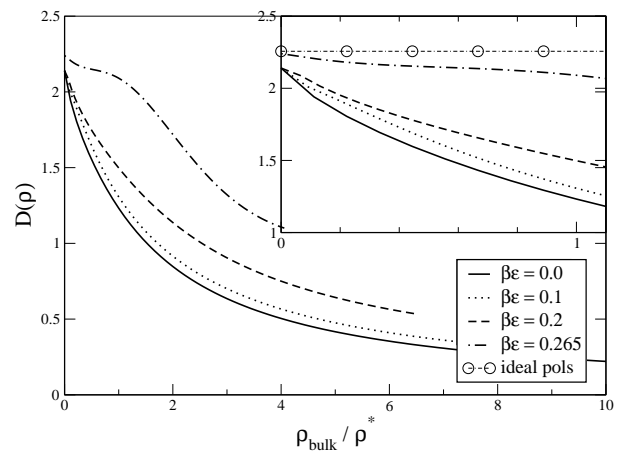}
\caption{ \label{fig:neatdep2} Range of the depletion potential
$D_w(\rho)$, as given by Eq.~(\protect\ref{eq2.6a}).  In the dilute
regime, highlighted in the inset, theta polymers are most
similar to ideal polymers, whereas polymers in good solvent already
show fairly strong deviations: at $\rho/\rho*=1$ the range has dropped
by almost a factor 2 compared to ideal polymers.  }
\end{figure}

In Fig.~\ref{fig:neatdep2}, we plot the range $D_w(\rho)$ for a number
of different solvent qualities.   In the dilute regime, highlighted
in the inset, the range for the near theta solvent is fairly close to
that of ideal polymers, whereas for polymers in good solvent the range
decreases markedly in the dilute regime.   In the semi-dilute regime
both good and theta solvent regimes show important deviations from 
ideal polymer behavior.  From scaling theory  we expect $D_w(\rho)
\sim \rho^{-0.77}$ and $D_w(\rho) \sim \rho^{-1}$ for the polymers
in good and theta solvents respectively.

\begin{figure}
\includegraphics[width=8cm]{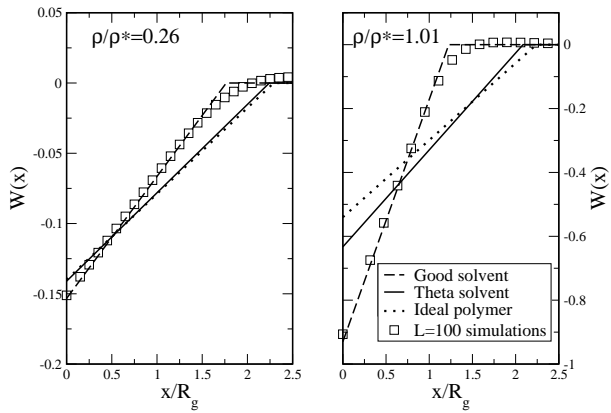}
\caption{ \label{fig:neatdeprho1} Depletion potentials, calculated via
  Eq.~(\protect\ref{eq2.6}), for $\rho/\rho*=0.26$ and
  $\rho/\rho*=1.01$.  This simple theory compares well to direct
  simulations, taken from \protect\cite{Loui02b}, for good solvents.
  For theta polymers the depletion potential resembles that of ideal
  polymers: at $\rho/\rho* = 0.26$ they are virtually
  inistinguishable; at $\rho/\rho*=1.01$ there are slight differences.
  }
\end{figure}

The results for $W(0)$ and $D_w(\rho)$ can be combined  with
Eq.~(\ref{eq2.6}) to determine the full depletion potential between
two plates. An example, for $\rho/\rho* \approx 1$, is shown in
Fig.~\ref{fig:neatdeprho1}.   The depletion potential for theta
polymers most closely resembles that of ideal polymers.  For lower
$\rho/\rho*$ we expect  an even closer correspondence.

In the dilute regime, Figs.~\ref{fig:neatW0}--\ref{fig:neatdeprho1}
suggest that the depletion potential between two plates, induced by
theta polymers, resembles that of ideal polymers.  We have recently
shown how to construct depletion potentials between two spheres from
that between two plates\cite{Loui02b}, and found quantitative
agreement with direct simulations of $L=500$ SAW polymers between
spheres, and also good results for ideal polymers.  This suggests that
the same procedure should work well for the depletion potential
between two spheres induced by theta polymers.  Furthermore, since the
depletion potentials are the dominant determinant of depletion induced
phase separation for $R_g/R_c \alt 1$\cite{Meij94,Rote04}, we would expect
the phase-behavior of theta polymers to closely follow that of ideal
polymers, at least for these size ratios.  If there are slight
deviations, then Fig.~\ref{fig:neatdeprho1} suggests that the binodals
would shift toward those of polymers in good solvent.  Recent
experiments\cite{Shah03} of colloids mixed with polymers in a near
theta solvent indeed show behavior that more closely resembles that
of ideal polymers than that of polymers in good solvent.

\section{Conclusions}

We have performed extensive Monte Carlo simulations to investigate the
bulk and interfacial properties of polymers in good solvent, and
polymers in the theta regime.  For infinite length $L$, there would be
a sharp transition between the behavior of polymers in good solvent,
and those in a theta solvent.  The simulations were carried out for
$L=500$, so that continuous cross-over effects are to be
expected\cite{Gras95,Shaf99}.  Working out the detailed crossover
behavior would require many more simulations at different
$L$. However, in practice, most experimentally investigated polymer
solutions do not reach the scaling regime either.  Moreover, we do
observe significant qualitative differences between theta polymers,
and polymers with weaker monomer-monomer attraction.

In contrast to polymers in good solvent, solutions of polymers in
theta solvent are quite well described by ideal polymer theories
throughout the entire dilute regime.  This works best for bulk
properties like the EOS, and slightly less well for interfacial
properties such as the density profiles, adsorptions, and surface
tension.  However, as predicted by theory\cite{deGe79,Rubi03}, in the
semi-dilute regime theta polymers also begin to to exhibit important
deviations from ideal-polymers.  Our simulations show behavior which
is consistent with that predicted by scaling theories, but the
concentration range we were able to investigate is not wide enough to
unambiguously confirm the expected scaling behavior.  We also suggest
that, for large enough concentrations, the values of various
properties, including the EOS, the adsorption, and the surface
tension, will be larger for theta polymers than for polymers in good
solvent at the same reduced concentration $\rho/\rho*$.

The close agreement between theories for ideal polymers, and our
simulations of theta polymers in the dilute regime, suggests that the
effective depletion pair potentials, and associated phase behavior,
should also resemble those of ideal polymers, at least in the
so-called ``colloid limit'', where $R_g/R_c \alt 1$.

\acknowledgements

We thank P.G. Bolhuis and V. Krakoviack for use of their computer
codes, and for their help, and Andrea Pelissetto for valuable
discussions concerning the scaling limit.  CIA thanks the EPSRC for a
quota studentship, and AAL thanks the Royal Society for their
financial support.


\begin{thebibliography}{99}
\bibitem{Log} Because $T_\theta$ is a tricritical point\cite{deGe79},
there should be logarithmic corrections to the scaling behavior.
 Renormalization group (RG) theory predicts\protect\cite{Shaf99} 
\begin{equation}\label{eq:log}\nonumber
R_g^2 \propto L\left( 1 -\frac{493}{5808} \ln^{-1}(L) + {\cal O}
\left( \frac{\ln(\ln(L))}{(\ln(L))^2}\right) \right),
\end{equation} 
but these corrections are of the
order of our simulation error bars, and so could not be resolved.
\bibitem{crossover} Note that for $L=\infty$, we would expect the EOS
to fall onto the same master curve for all temperatures $T > T_\theta$
in the good solvent regime. The fact that they do not in
Figs.~\protect\ref{fig:neatopress}~and~\protect\ref{fig:neatlogplot}
is most likely due to finite $L$ crossover effects, which are also
evident in some of the other physical properties studied in the
current paper.
\end{thebibliography}
\end{document}